\documentclass[11pt]{asaproc}
\usepackage{caption}
\usepackage{graphicx}

\usepackage{times}

\title{ Satellite Images and Deep Learning to Identify Discrepancy in Mailing Addresses with Applications to Census 2020 in Houston}

\author{Zhaozhuo Xu\thanks{Department of Computer Science, Rice University}  \and  Alan Baonan Ji\footnotemark[1] \and  Andrew Woods\thanks{Department of Political Science, Rice University} \and Beidi Chen\thanks{Computer Science Department, Stanford University} \and Anshumali Shrivastava\footnotemark[1]}
\begin{document}

\maketitle

\begin{abstract}
The accuracy and completeness of population estimation would significantly impact the allocation of public resources. However, the current census paradigm experiences a non-negligible level of under-counting. Existing solutions to this problem by the Census Bureau is to increase canvassing efforts, which leads to expensive and inefficient usage of human resources. In this work, we argue that the existence of hidden multi-family households is a significant cause of under-counting. Accordingly, we introduce a low-cost but high-accuracy method that combines satellite imagery and deep learning technologies to identify hidden multi-family (HMF) households. With comprehensive knowledge of the HMF households, the efficiency and effectiveness of the decennial census could be vastly improved. An extensive experiment demonstrates that our approach can discover over 1800 undetected HMF in a single zipcode of the Houston area.

\end{abstract}






\section{Introduction\label{intro}}
The accuracy and completeness of the decennial census are required in the allocation of federal funds, grants, and support to states, counties, and communities so that they can ensure communities receive the needed resources for a successful census campaign~\cite{salling2020we}. Furthermore, many businesses rely on census results to determine locations to build factories, offices, and stores, which creates local job opportunities. However, the US's census, although having experienced progress in the past decades, still suffers from a non-negligible level of under-counting. Particularly, due to the scarcity of human labor resources, the census is unable to cover all regions for a large city such as Houston. According to the latest statistics, ~\cite{undercount}, the average tract-level response rate among all census tracts over the US is $61.5\%$. For tracts with a very high risk of under-counting young children, this rate is only $55.3\%$. These relatively low response rates represent a significant degree of information loss in specific neighborhoods, which leads to an unbalanced allocation of public resources. 
\begin{figure}[t]
    \centering
\centering
\includegraphics[width=1\textwidth]{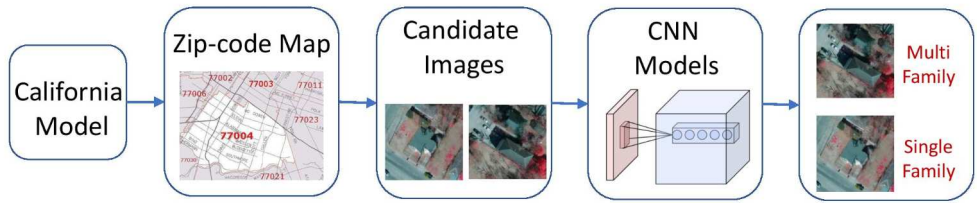}
\captionof{figure}{Method overview}
 \label{fig:overview}
\end{figure}

The existing solution to this problem by the Census Bureau is to increase canvassing efforts by recruiting more canvassers or volunteers~\cite{kissam2018community} for door-to-door interviewing, mail-in survey, and street canvassing. However, this approach suffers from two practical problems. The first problem is that it is not guaranteed that sending more canvassers will lead to more accurate counting. Whether canvassers can alleviate the underestimation is highly dependent on the location of surveying efforts. For example, canvassers surveying an area with an originally high response rate for the past census is much less impactful than canvassers surveying potential areas with low response rates. Additionally, several governors in the local Census Bureau vote and determine a priority list of zip code regions to send canvassers. These lists are based on various regional statistics, which include income group, ethnicity, response rate, and degree of under-counting in the past years. This procedure has two major disadvantages: (1) last year's statistics, especially response rate and degree under-counting, may be outdated due to rapid city development, and (2) the voters' inputs and experiences may not be accurate, which heavily affects the region selection. The second problem is that the most important survey method, the mail-in survey, has an even higher risk for under-counting. This canvassing paradigm is that every valid USPS address gets a form, and the census collects population information based on returned forms. But, for instance, a recently converted multi-family unit may only be sent one such form. If the canvassers have little knowledge of that unit's updated state, the forms will be sent and collected with blind knowledge about the multi-family situation for that household.

In this paper, we closely analyze the inaccurate and costly census paradigm and argue that the existence of hidden multi-family households is a major cause for under-counting. Furthermore, we can largely reduce under-counting if we can have a low-cost method to identify all hidden multi-family (HMF) households. However, HMF identification is still challenging. First, the brute force way of in-person identification remains costly. An alternative method could be using supplementary information such as electricity or water usage per household to infer HMF. However, privacy issues are of concern since this information is often prevented from being publicly accessible. Furthermore, it is hard to handcraft features that relate to complex HMF patterns from using statistical methods for predictions.

In contrast, we propose a novel formulation that consists of two steps: (1) Identifying the hidden multi-family households via a satellite-based detector. (2) Given the knowledge on the addresses of multi-family households, allocate the canvassers to minimize undercounting.  We take advantage of satellite imagery's public availability and the abundant information that it contains. Combined with advanced deep neural networks with knowledge transferred from public benchmarks~\cite{imagenet_cvpr09}, we learn to identify HMFs from end to end without in-site canvassing or handcraft feature engineering. Our method leads to two major benefits: (1) Intelligent region selection: with missed addresses retrieved by our approach, we are able to allocate the concentration of the campaign into the area with the most under-counted regions. (2) Smart canvassing: given the marked latitude and longitude of hidden multi-family addresses, canvassers have a direct coordinate during the field investigation instead of randomly sampling addresses, which largely increases the census efficiency. 
After extensive experiments, we demonstrate that our approach can discover over 1800 undetected HMF in a single zipcode of the Houston area.

\section{Related Works}

Satellite imagery has been widely used in traditional tasks such as high-resolution land-use classification~\cite{yang2019dynamic} and small object recognition~\cite{xu2017deformable}. However, the remote sensing technologies in social-economic tasks such as poverty mapping~\cite{jean2016combining} remain at the district level. For census,  \cite{harvey2002estimating} proposes a method that combines the satellite imagery as a calibration for district-level population counts. \cite{wardrop2018spatially} also includes satellite imagery as an aggregation resource for area-level population counts. Using satellite imagery for a much more detailed characterization of households remains a challenging task. Moreover, these approaches treat satellite imagery and population counts independently, and satellite imagery cannot provide productive guidance during the census process. Therefore, combining satellite imagery and census research in a more interactive way is a promising but challenging direction for research.



\section{Methods}
\subsection{Overview}
In this section, we present the overview of our approach. As shown in Figure~\ref{fig:overview}, we present the workflow of our method. Our approach contains three major parts: Effective Area Identification and  Data Collection and Detector Training. We first identify the area that will be most effective based on California Model~\cite{kissam2018community}, and then train a satellite image-based detector to collect both hidden and marked multi-family households. Combining with existing single-family households, we train and generate a satellite-based HMF detector, which can guide canvassers in finding the locations of HMF.


\subsection{Effective Area Identification}
To begin developing the model, we calculated - for each tract in Harris County - an estimate of the likelihood
of a tract containing an irregular housing unit. This scoring method was based on previous research aimed at
identifying irregular housing units in California communities~\cite{kissam2018community}. Since we were unsure whether this method
would accurately identify these households in a context outside of California, we relied on canvasser input to
validate specific tracts. In particular, from the list of all tracts, we designated around 200 for canvassing
according to both a Bad MAF Score, which is defined, based on the California model, as a tract's overall degree of under-counting, as well as the Low Response Score provided by the Census Bureau. After receiving
initial feedback from canvassers, we used the modified list as a guide for selecting areas for further investigation using the satellite imagery model.  According to Table ~\ref{table:area}, we suggested 50 $\%$ of canvassers' efforts on the high Bad MAF Score (overall degree and high Low Response Score. After further ranking the regions, we choose zipcode 77004 as our top priority candidate for further study. 

\begin{table}
\caption{Identification of Effective Area by California Model }
\begin{center}
\begin{tabular}{ccc}
\hline
\hline
\\[-5pt]
Bad MAF Score & Low Response Score & Suggested Canvassing Efforts\\
\hline
High&     Low&    50$\%$\\
High&     High&      20$\%$\\
Medium&     Low&   15$\%$\\
High&     High&       15$\%$\\
\hline
\end{tabular}\label{table:area}
\end{center}
\end{table}

\subsection{Data Collection}\label{sec:data_collect}
Our proposed model uses three main data sources. First, we used the Aerial Imagery 2018 dataset from the
Kinder Institute, which contains 700 image tiles with a map scale of 1 "=100' with a 6-inch ground sample
distance (GSD) for Harris County. We also used the Harris County Appraisal District (HCAD) Advanced
Records, which provided advanced property search metrics to identify multi-family addresses within a specified zipcode, and B1 State Category (residential, multi-family). Finally, we used the HCAD Parcels and Address Points data to generate our single-family addresses. In order for the model to properly identify multi-family housing units on its own, we must first train the model to identify such housing units from a base list of known multi- and single-family units. To do this, we relied on the data resources listed above as well as the Google Maps API to retrieve latitude/longitude coordinates for each single and multi-family address we observe in the address records data. We were able to finalize a satellite imagery dataset of ~300 multi-family and ~10,000 single-family households. In our initial assessment, this dataset only covered the zip code 77004 and contained corresponding image tiles per category (multi-family vs. single-family). The bounding boxes for each satellite image are 50 by 50 meters around each
house, with each center location derived from the HGAD address points data. 
\subsection{HMF Detector}

\subsubsection{Hypothesis}
Since irregular housing is an umbrella term for a variety of potential housing arrangements, we initially limited our analysis to a particular type of housing arrangement - multi-family households. A multi-family household is a single traditional housing structure that has been sub-divided to accommodate multiple families. These structures are different from apartment complexes or townhouses, which feature a unique housing structure featuring originally planned and repeated divisions. Multi-family homes, on the other hand, often feature unplanned and ad hoc divisions such as the addition of an interior wall, an appended structure, or a fence designed to separate the multiple households that live in what is technically the same building.

Following this direction, our work starts with a major hypothesis: deep learning-based algorithms can identify HMF from satellite image tiles.  Satellite images have proved to be an efficient way of poverty detection[1] or other social or economic tasks. However, as the application remains at the area level, the satellite-based learning algorithm's performance in household-level tasks remains undiscovered.  In this work, we investigate satellite-based learning in predicting the properties of individual houses.  Our goal is to guide canvassers by detecting HMFs in satellite imagery.

\subsubsection{Neural Network Training}

To tackle the problem of HMF houses detection, we first developed a Smart Data Collection approach that was used to identify an Effective Area of study that would contain a good mix of potentially concealed, multi-family households with single-family households. To generate candidates for the Effective Area, we utilized the pilot model to rank zipcodes within Harris County. Once we finalized the zipcode for further study, we collected both multi-family and single-family addresses from HCAD Advanced Records. We cropped 50m x 50m image tiles from satellite imagery to train our network. We began the training process
by splitting the dataset into training, validation, and test sets, for which we used three different model types- ResNet-101 with ImageNet pre-trained weights, VGG-16, and DenseNet-121 to train our model. According to our assumption, DenseNet\cite{huang2017densely} may perform the best, with ResNet and VGG following. Afterward, equipped with the deep detector from satellite imagery, we were able to retrieve many ranked potential multi-family houses within our effective 77004 zipcode, and as a result, could be inferred over the entire Harris County, with much more detected tiles. We ranked these tiles based on multi-family resemblance.

From our initial investigation, we were able to extract a number of satellite images that give us a sense of
how the model is delineating multi-family from single-family households. For example, we obtain a satellite image of households labeled as single-family by the HCAD but detected as multi-family by our model. As noted before, these single-family units encompass both traditional single-family houses as well as townhouses and apartment complexes - something clearly reflected in the images and the model predictions more broadly.



\section{Experiments}
\subsection{Objectives}
The objective of the experimental evaluation is to investigate the performance of our satellite image-based hidden multi-family house detector in real scenarios. The performance includes both the accuracy in established datasets as well as in-field discovery. Specifically, we target to answer the following questions:

\begin{itemize}
    \item Do deep convolutional neural networks achieve satisfactory results in distinguishing hidden multi-family households from single-family households?
    \item  Among different deep neural networks architecture, which one achieves the best performance in our task?
    \item  Outside from the established datasets, what is the performance of our detector on real-world scenarios?
\end{itemize}

\begin{figure}
\centering
\includegraphics[width=.8\textwidth]{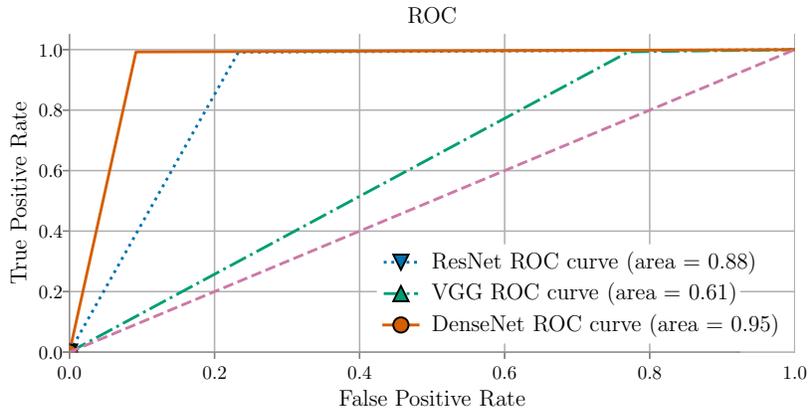}
\captionof{figure}{ROC curves of three deep neural networks. We show that DenseNet outperforms ResNet and VGG with better ROC curves.}
 \label{fig:roc}
\end{figure}
\subsection{Settings}
\label{sec:setting}
\subsubsection{Datasets}
We generate a dataset for HMF detection. As shown in Section~\ref{sec:data_collect}, we collected a set of satellite image tiles and combined it with the web crawled binary class labels from HCAD Advanced Records, which determine whether households are single- or multi-family.  Our dataset contains a total of 2800 tiles, including 280 multi-family tiles. We first split the dataset as train, validation, and test by a ratio of  0.64:0.16:0.2. Afterward, we tuned the parameters on the training dataset to maximize the accuracy of the validation dataset. Finally, the accuracy of the test dataset will be presented for comparison.

\subsubsection{Experiment Settings}
We implemented all deep networks using Keras~\cite{gulli2017deep}. The preprocessing of satellite imagery is based on python with ArcGIS API~\cite{johnston2001using}. All the experiments were conducted on a machine equipped with two 20-core/40-thread processors (Intel Xeon(R) E5-2698 v4 2.20GHz) and one NVIDIA Tesla
V100 Volta 32GB GPU. The server has an Ubuntu 16.04.5
LTS system with the installation of Tensorflow-GPU as the backend for Keras.

\subsubsection{Evaluation Metrics}

For evaluation of accuracy in our dataset, we introduced the ROC curve~\cite{mcclish1989analyzing} with AUC score~\cite{rosset2004model}. We also included confusion matrices for model comparison. 

\subsection{Results on Established Dataset}

\subsubsection{Main Results}
In this section, we presented the ROC curves along with AUC scores of three deep neural networks used for multi-family detection. From Figure~\ref{fig:roc}, we observe that deep neural networks achieve over 0.6 AUC on the HMFs classification. The maximum AUC is 0.95, which is high enough for distinguishing HMF households from single-family households. These results answer the first proposed question: deep convolutional neural networks are capable of classifying multi-family households from single-family households, indicating its potential in guiding the canvassers to reduce the under-counting effect in the census. 

\begin{figure*}[t!]
	\begin{center}
		\begin{tabular}{ccc}
		\hspace{-1cm}
			\includegraphics[width=0.4 \textwidth]{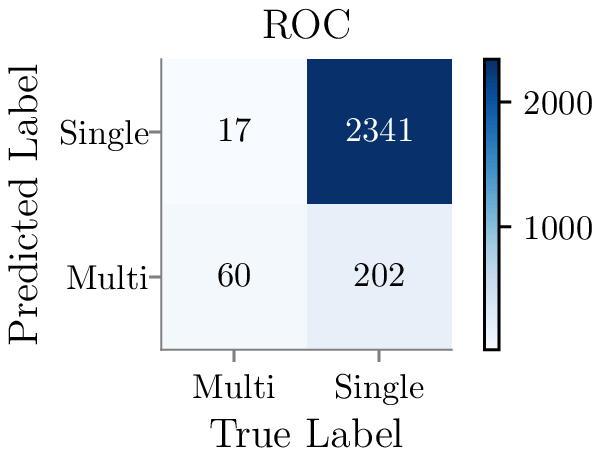} &
			\hspace{-1cm}
			\includegraphics[width=0.4 \textwidth]{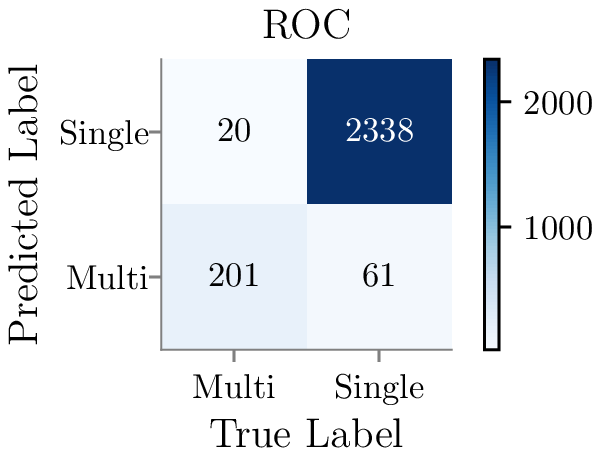} &
			\hspace{-1cm}
			\includegraphics[width=0.4 \textwidth]{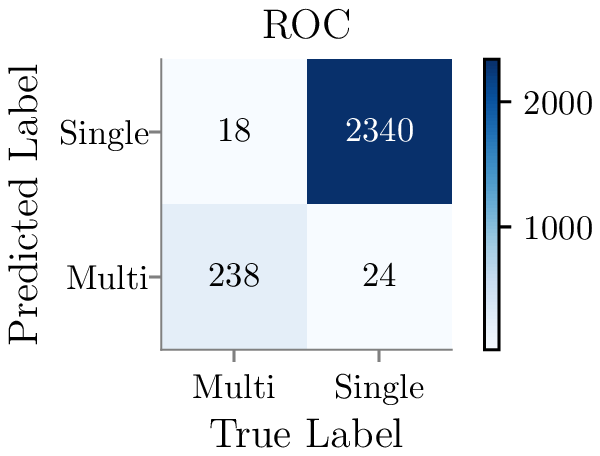}
		\end{tabular}
	\end{center}
	\caption{Confusion matrices of three deep neural networks. Left: VGG, Middle: ResNet, Right: DenseNet.  The confusion matrices validate that DenseNet has better performance in HMF detection.}\label{fig:confuse}
\end{figure*}

\subsubsection{Model Comparison}

In this section, we provide a comparison of three introduced deep models in our task. In Figure~\ref{fig:roc} and Figure~\ref{fig:confuse}, we present the comparison in the form of ROC curves and confusion matrices. According to the demonstration, DenseNet achieves better performance over ResNet and VGG networks. Specifically, the DenseNet hash has fewer false positives and true negatives, indicating its superiority in our task.  This observation answers the second question: DenseNet should be our choice for the HMF households detection for its advantages over other deep neural networks empirically.

\subsection{Results on Real World Discovery}

\begin{figure*}[t!]
	\begin{center}
		\begin{tabular}{cc}
		\hspace{-0.5cm}
			\includegraphics[width=0.2925\textwidth]{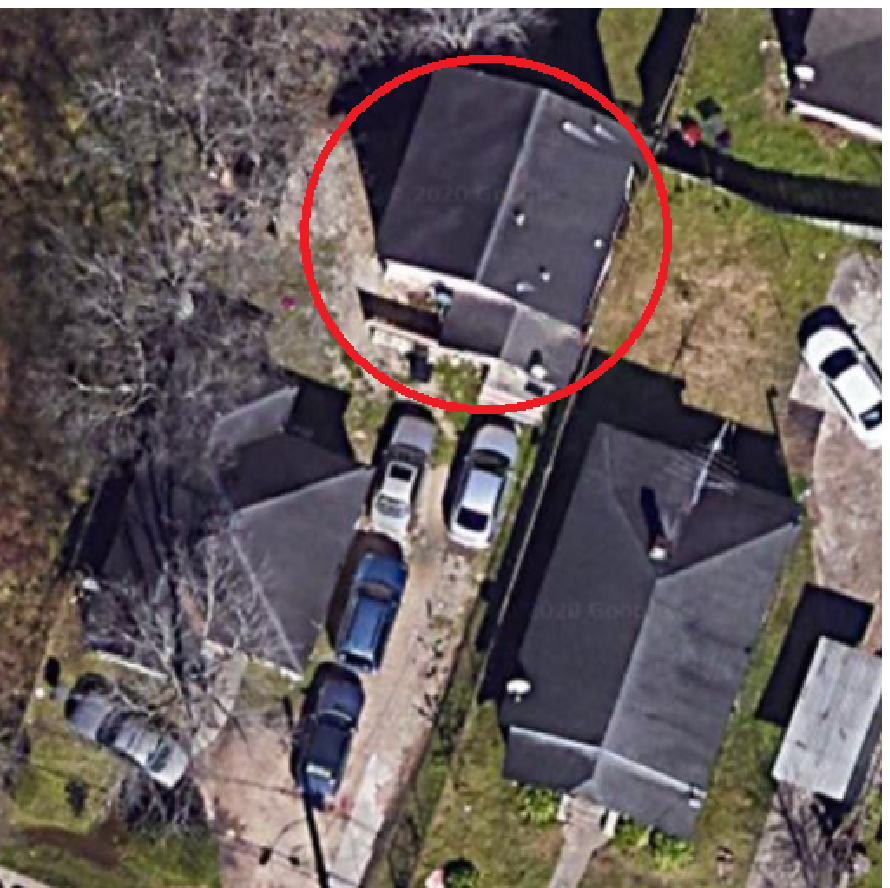} &
			\hspace{-0.5cm}
			\includegraphics[width=0.35\textwidth]{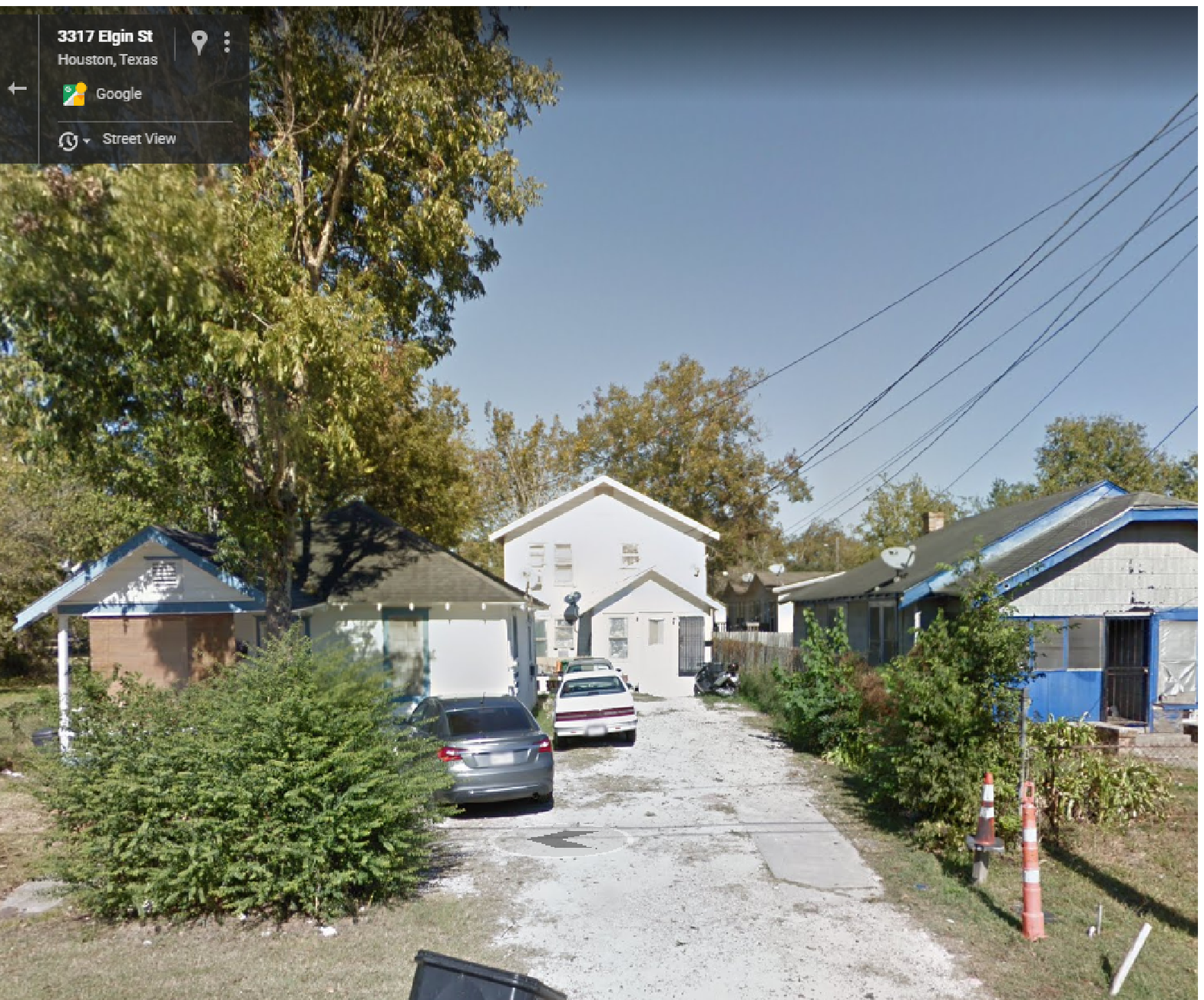} \\
			\hspace{-0.5cm}
			\includegraphics[width=0.297\textwidth]{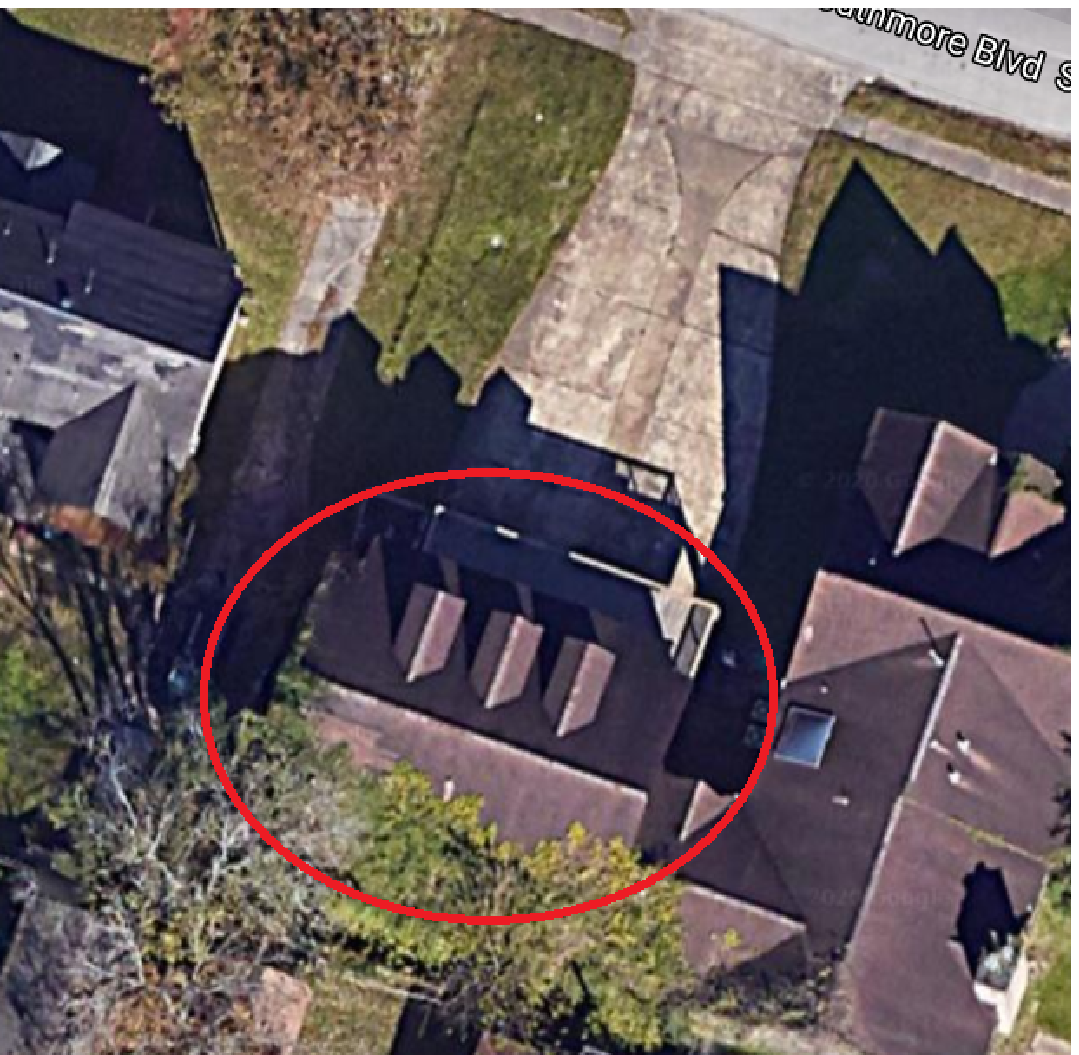} &
			\hspace{-0.5cm}
			\includegraphics[width=0.35\textwidth]{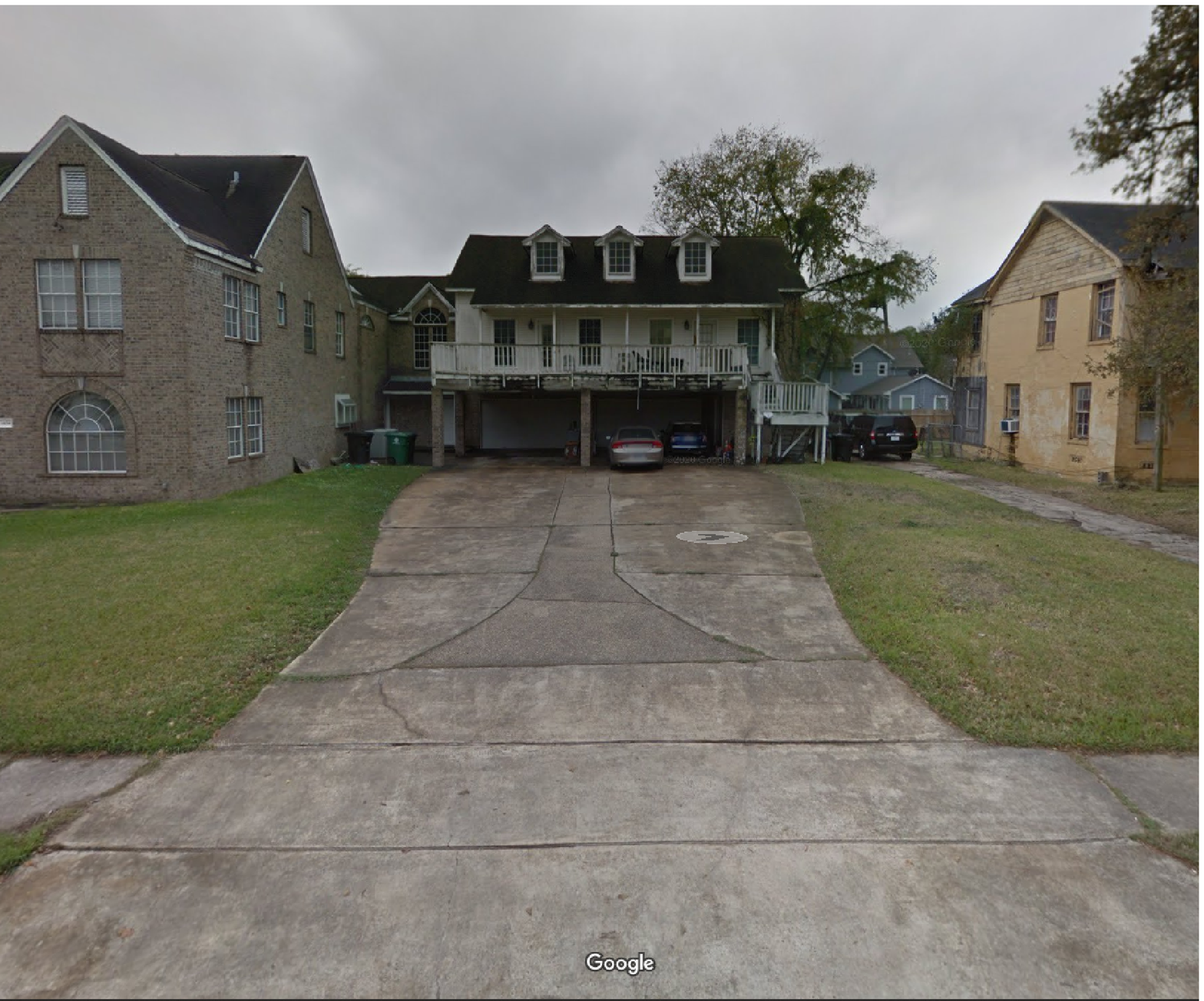}
		\end{tabular}
	\end{center}
	\caption{Satellite view versus street view of detected hidden multi-family houses. Left: The satellite view of two miss-labeled multi-family households. Right: their corresponding street view.}\label{fig:satelite}
\end{figure*}

In this section, we present the results of our methods in real-world scenarios. To test the generality of our detector, we run model inference on all tiles marked by the HCAD in a single zipcode area. The objectives of this procedure are: (1) Validate the performance of detector on unseen image tiles (2) Perform real-world discovery on the Houston area. As a result, we extract over 1800 suspect HMF households from that zipcode area. In Figure~\ref{fig:satelite}, we provide an example of the detected hidden multi-family household. On the left, we present the satellite view of the address. Compared to the street view on the right. It is clear that this address is a multi-family household. However, in HCAD system, this address is marked as single-family. After collaboration with the census organization, we conclude that the detected households are at high risk of being under-counted during the regular census campaign. 

\section{Broader Impact}
Overall, our machine learning model shows promise as a viable method to support Census operations but
is still limited in its applicability and efficiency due to the challenges posed by the pandemic. Aside from
our inability to send canvassers to directly investigate tracts and households of interest, another major limitation our approach has faced is the lack of access to the Census Bureau's Master Address File (or a
similar document). While the HCAD data and other sources are helpful, they do not represent the "final
word" on whether an address will receive a Census for, nor do they necessarily indicate that a particular
household is fully known to the Bureau. To move forward with this method, a more complete address list
would be critical to reducing misclassification and maximizing resource allocation.

These challenges aside, this methodology has potential application to a broad array of challenges beyond
counting in the decennial census. For instance, a similar methodology could be developed in order to track the development of new housing units across a geographic area and note population trends across time. Similarly, efforts to identify multi-family housing could be especially useful during times of displacement due to natural disasters like hurricanes. Finally, efforts to identify concealed housing, in particular, could also be used by communities and organizations to locate potentially unregistered or inactive voters or simply to better understand community needs and identify at-risk individuals. All of these applications are possible using the current or a similar model conditional on resources being devoted to the validation of the model predictions and the iterative development of model accuracy. They highlight the great promise of using satellite imagery and machine learning to support activities around the Houston area.

\section{Conclusion}
The accurate and complete detection of hidden multi-family (HMF) houses plays an essential role in discovering the underestimated population for Houston Census 2020.  The current way of HMF detection relies on canvassers investigating individual addresses physically. Due to limited human resources, this procedure is ineffective and inefficient.  In this work, we observe the HMF in a satellite view. We propose an automatic algorithm that identifies HMF from satellite image tiles.  An extensive experiment demonstrates that our approach can discover over 1800 undetected HMF in a single Houston-area zipcode.

\bibliographystyle{unsrt}
\bibliography{ref}

\end{document}